\documentclass[10pt,twocolumn,aps,prb,twoside]{revtex4}
\pdfoutput=1

\usepackage{graphicx}        
\usepackage[T1]{fontenc}
\usepackage{mathptmx,courier,helvet}
\usepackage{textcomp}

\usepackage{fancyhdr}
\fancypagestyle{plain}{%
        \fancyhead[LE,RO]{}
        \fancyhead[LO,Re]{}
          \fancyfoot[C]{Draft, Dec 2012, arXiv.org.}
           }
\graphicspath{{pic/}}

\usepackage{amsmath}

\usepackage{listings}

\begin{document}

\title{Device model of silicon nanowire bioFETs}

\author{M.~W.~Denhoff} 
\affiliation{M.~W.~Denhoff is with the Institute for Microstructural Sciences
of the National Research Council of Canada, Ottawa, ON, K1A 0R6, Canada.}
\email{mike.denhoff@nrc.ca.}
\date[Draft posted to:]{arXiv.org, Dec 2012}

\begin{abstract}
A device model of biological molecule sensors based on semiconductor nanowires
has been developed.
This model of a bioFET is based on the concept of the electrolytic
absolute electrode potential.
From that starting point a semiconductor device model of the nanowire solution
biomolecule system was derived.
The model includes the Gouy-Chapman-Stern model of the salt solution double layer,
site binding charges on the electrode surface, and biological molecules in
the form of a membrane layer.
A simple method of solving this model is presented using the finite element
method.
Some examples showing the general properties of the model are given.
\end{abstract}

\maketitle
\thispagestyle{plain}

\section{Introduction}

The idea of using a field effect transistor (FET)
to detect charged  molecules in a solution
was first introduced by Bergveld.\cite{bergveld,bousse1}
Since then there has been much research into using FET based devices
to detect charged biological molecules (bioFET).\cite{schoning1,schoning2}
In the past 10 years or so research has turned to nanowire FETs,
with a number of workers reporting high sensitivities,
\cite{cui,stern,ingebrandt,lu} but the
results are still variable and not well understood.\cite{schoning2}

In order to understand the experimental results, there is a 
need for a simple, accurate, and easy to use model.
The basic principle of bioFET operation is that charged ions
attached to the gate oxide will attract or repel carriers in the
FET channel, changing the channel conductivity.
A model needs to describe the structure of the mobile ions
in the electrolyte, surface charging of the gate insulator,
along with the biomolecules and how these interact with the
carriers in the semiconductor.

A number of
analytic treatments have been published.\cite{bousse1,gao}
These use approximations to render the mathematics solvable,
and are necessarily incomplete,
 so only apply to restricted cases.
The complexity of the problem leads to using numerical methods.
Early numerical models were one dimensional and divided the system into
a number of layers; essentially a series of capacitance's.
Poisson's equation was solved in each layer 
in sequence while matching boundary conditions.
This process was then iterated until an over all consistent
solution was obtained.\cite{sandifer,landheer}
This is a somewhat awkward and complex procedure that has only
been applied to planar structures.

Semiconductor style modeling is the next step in
numerical modeling.\cite{li,nair,liu,baumgartner,chung,vico}
This merging of semiconductor and electrolyte regions
 is a complex system, leading to examples of
incompletely explained or even incorrect models.
Recently, Dutton published results of a fairly complete
numerical device model.\cite{liu}
Even this publication does not explain the basis of the model,
boundary conditions, or how they performed the calculations.
Without these details, it is difficult for other workers to
 try to duplicate their results.
A very detailed model  solves the full semiconductor
model in the Si, including the continuity equations and a separate
Monte Carlo simulation of the charge in the
layer of biomolecules.\cite{baumgartner}
This model is complicated and the calculations are difficult.
Since the semiconductor part and the biomolecule part are separated,
the entire solution may not be self consistant.

Other recent work\cite{chung} uses proprietary software to solve the
semiconductor equations in 3 dimensions.
This allows simulation of incomplete coverage of a bioFET using charged
cubes to represent biomolecules.
This model, however, does not fully represent a bioFET as it does not
include features such as the Stern layer and specific surface 
charges (eg. site-binding model).

In many experimental situations it can be arranged that the bioFET
surface is uniformly coated with biomolecules.
This case can be modeled as a 2 dimensional membrane, which will
be done in this paper.
It would seem expedient to concentrate on this simpler situation,
at least until basic bioFET behaviour is understood.
Then one could move to the more complicated case of partial
coverage.\cite{chung}

The purpose of this paper is to give a basic bioFET
device model of a semiconductor-interface-electrolyte system
(Si/SiO$_2$/Sol) and a simple method to calculate the results.
This model should be based on semiconductor physics and electrochemical
principles so that it can serve as a basis for further development.
The idea is to include all the important features of the bioFET
system and using as simple a calculation as possible.
This is to allow easy comparison between theory and experiment
to facilitate investation of the properties of bioFET devices.

The semiconductor-interface part will be represented by
 Si with a thin SiO$_2$ surface layer and the electrolyte as a
 salt solution.
The next section will discuss the electrolytic cell in terms
of absolute electrode potentials.
In the following section the view will change from the
electrochemical view to the semiconductor device view.
The device model will
include a band structure for the Si/SiO$_2$/Sol
system, electron and hole densities in the Si and the structure of
ion concentrations in the solution.
A membrane model will be used to model a charged biomolecule
layer.

A finite element method of calculation will be outlined.
The results for some basic examples will be
 used to demonstrate 
 cylindrical nanowire end conditions, and
calculation of response or sensitivity.

\section{Electrochemical cell}

While this paper is concerned with modeling the Si/SiO$_2$/Sol
system, an experimental system must consist of a complete electrolytic
cell with two electrodes.
The concept of absolute electrode potential will be used here
following the detailed work of Trasatti and
 Reiss.\cite{trasatti89, reiss, trasatti80} 
Consider the cell
\begin{equation}
\mathrm{M}' \overset{1}{\big |} \mathrm{Si} \overset{2}{\big |}
    \mathrm{Sol}
   \overset{3}{\big |} \mathrm{M} 
\end{equation} 
The Si$\big|$Sol interface represents the system that we wish to model.
The thin SiO$_2$ layer is omitted here for simplicity.
The Sol$\big|$M junction represents a reference electrode.
A complete circuit is formed by connecting the reference electrode, M,
to an ohmic contact on the Si electrode with a wire made
of the same metal as the reference electrode, M$'$.
An ohmic contact means that M$'$ and Si are in electronic equilibrium,
thus the electrochemical potentials in M$'$ and the Si are equal.
The full cell potential can be written as the difference of the
absolute electrode potentials,
\begin{equation}\label{eq:electrode}
E = E^\mathrm{Si}_\mathrm{abs} - E^\mathrm{M}_{abs}.
\end{equation}
There are a number of possible definitions of absolute electrode potential.
The definition that is useful here
uses a ``free'' electron in the solution as the reference state,
Trasatti's $_1E^\mathrm{M}_\mathrm{a}$.\cite{trasatti89}
For the Si electrode, this gives
\begin{equation}\label{eq:absSiSol}
E^\mathrm{Si}_\mathrm{abs} =
   -\left(\frac{\overline\mu^\mathrm{Si}_\mathrm{e}}{q} -
      \frac{\overline\mu^\mathrm{Sol}_\mathrm{e}}{q}\right),
\end{equation}
where $\overline\mu^\mathrm{Si}_\mathrm{e}$
 and $\overline\mu^\mathrm{Sol}_\mathrm{e}$
are the electrochemical potentials of electrons in the Si and the solution,
respectively.
Substituting the definition of electrochemical potential gives
\begin{equation}
E^\mathrm{Si}_\mathrm{abs} =
  -(\frac{\mu^\mathrm{Si}_\mathrm{e}}{q}-\phi^\mathrm{Si}
    -\frac{\mu^\mathrm{Sol}_\mathrm{e}}{q}+\phi^\mathrm{Sol}),
\end{equation}
where $\mu^\mathrm{Si}_\mathrm{e}$
 and $\mu^\mathrm{Sol}_\mathrm{e}$
are the chemical potentials of electrons in the Si and the solution,
respectively,
 $\phi^\mathrm{Si}$ is the inner electrostatic potential
in the Si bulk and $\phi^\mathrm{Sol}$ is the inner
 electrostatic potential in the solution bulk.
$q$ is the elemental charge.
The inner potential is given by $\phi=\chi+\psi$,
where $\chi$ is the surface polarization and $\psi$ is the
outer electrostatic potential.
$\psi$ is generated by free charges at the surface of a phase
plus external field sources.
The electron work function of a material, $\Phi$, is the 
negative of the electron real potential, $\alpha_\mathrm{e}$, given by
$\Phi=-\alpha_\mathrm{e}=
     -\mu_\mathrm{e}+q \chi.$
Now (\ref{eq:absSiSol}) can be written in terms of measurable quantities,
\begin{equation}\label{eq:absSi}
E^\mathrm{Si}_\mathrm{abs} =
  \frac{\Phi^\mathrm{Si}}{q}-\frac{\Phi^\mathrm{Sol}}{q}
  + \psi^\mathrm{Si} - \psi^\mathrm{Sol}.
\end{equation}
In the case of electronic equilibrium for the Si$\big|$Sol electrode,
$E^\mathrm{Si}_\mathrm{abs}=0$ and 
\begin{equation}\label{eq:builtin}
\Delta \psi = \psi^\mathrm{Si} - \psi^\mathrm{Sol} =
     -\frac{\Phi^\mathrm{Si}}{q}+\frac{\Phi^\mathrm{Sol}}{q}.
\end{equation}
This gives the boundary condition for the electrostatic potential.
For a detailed picture of the inner potential in the Si/SiO$_2$/Sol
system, including the polarization at the interfaces see the paper
by Bousse.\cite{bousse2}

The voltage on the Si with respect to the solution, $V=E^\mathrm{Si}_\mathrm{abs}$,
can be expressed using (\ref{eq:electrode}) as $V = E^\mathrm{M}_\mathrm{abs} + V_\mathrm{app}$,
where $V_\mathrm{app}$ is applied between the electrodes by an
external power supply.

The approach to this problem used by Ref.~\onlinecite{chung} is to
ignore the chemical potential of the reference electrode, ie.
to assume an ohmic contact to the solution.
Then, they moldel the solution as a semiconductor with a 1.1\,eV
energy gap and with an artificially defined electron affinity.
The value of which is determined by matching their model
calculations to experimental data.

\section{Device model}

  The model of the system Si/SiO$_2$/Sol 
 will start by considering the equilibrium state
and then be extended to non-equilibrium with essentially zero current.
It is important to note that,
it is common, in semiconductor modeling, to use the outer potential, $\psi$,
rather than the inner potential, $\phi$.
$\psi$ does not include the surface and interface polarization and so
is continuous at interfaces.
The effects of surface polarization, $\chi$, is, arbitrarily, included in the
value of the work functions.

In the semiconductor device model one must solve the Poisson
equation,\cite{sze,kurata}
\begin{equation}\label{eq:poisson}
\nabla ( \epsilon \nabla \psi) = \rho.
\end{equation}
$\epsilon$ is the position dependent dielectric constant and 
$\psi$ is the electric potential.
$\rho$ is the charge density and is composed of electron, $e$,
hole, $h$, and dopant densities in the Si and of
ions in the solution.

In the general, non-equilibrium case, continuity equations
for each mobile particle must also be solved.
In the model presented here a voltage applied to the Si
with respect to the solution will result in
essentially zero current due to the insulating SiO$_2$ layer.
With zero current, the particle conservation equations are trivial
and the applied voltage can be accounted for in the expressions for
the charge densities.

In the next section the band structure of the system will be described,
followed by a development of the charge carrier statistics.
Then the structure in the electrolyte will be presented.

\subsection{Band structure}

\begin{figure}
\centering
\includegraphics{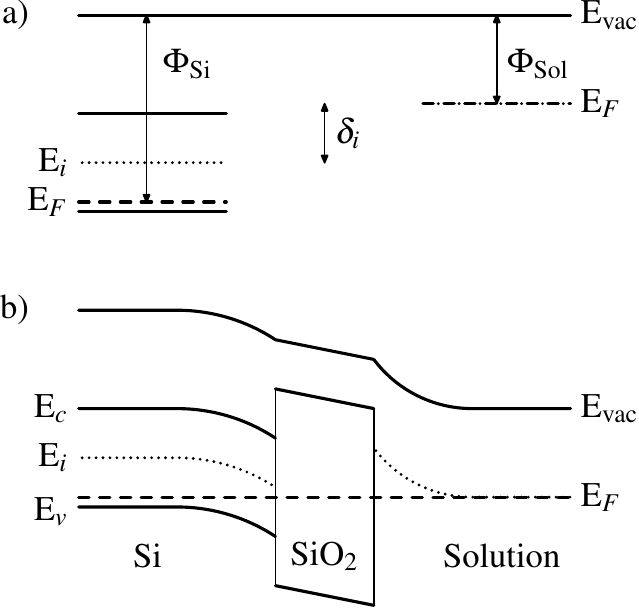}
\caption{Energy diagrams. a) Isolated p-type Si and solution phases showing
the work functions, $\Phi^\mathrm{Si}$ and $\Phi^\mathrm{Sol}$,
as well as the Fermi levels, E$_F$, and intrinsic levels, E$_i$.
In the solution, the intrinsic level
 is equal to the Fermi energy.
b) Si-SiO$_2$-electrolyte system at equilibrium.  The charge redistribution
generates an electric field 
which changes the energy of the
 vacuum level, E$_\mathrm{vac}$, and the bands.}%
\label{fig:energy}
\end{figure}

The Si/SiO$_2$/Sol system will be treated as a
semiconductor heterojunction system.\cite{sze}
  The solution can be thought of as
a low carrier density metal or a very small band gap semiconductor.
Consider isolated p-type Si and solution phases as drawn in
 Fig.~\ref{fig:energy}a).
The Fermi level in the solution coincides with the energy of an
electron in the neutral solution.
The work function for each material is the energy difference between
the Fermi level and the vacuum level.

Fig.~\ref{fig:energy}b) shows a band diagram where
 the system is at electronic equilibrium 
so that the Fermi level is constant across the system.
  To achieve this,
charge must be redistributed such that the resulting electrostatic
potential, $\psi$, bends the vacuum energy level by an amount equal to
the difference of the work functions.
This is the builtin potential and is expressed by (\ref{eq:builtin}).

When a voltage, $V$, is applied to the Si, the potential, $\psi$, is increased
by $V$ and the electron energy bands and the Fermi level are lowered
by $-V$.  The difference in the bulk values of $\psi$ in (\ref{eq:builtin})
is modified to 
\begin{equation}\label{eq:bc}
 \Delta\psi=-\frac{\Phi^\mathrm{Si}}{q}+\frac{\Phi^\mathrm{Sol}}{q}+V.
\end{equation}
The work function of neutral Si depends on the dopant density,
and the intrinsic energy.
It can be expressed by,\cite{sze}
\begin{subequations}\label{eq:wf}
\begin{align}
-\Phi^\mathrm{Si} = \mathrm{E}_F = \mathrm{E}_i + kT \ln\left(\frac{N_D}{n_i}\right),
          \ \ \ \ \mathrm{n-type}\\
-\Phi^\mathrm{Si} = \mathrm{E}_F = \mathrm{E}_i - kT \ln\left(\frac{N_A}{n_i}\right),
          \ \ \ \ \mathrm{p-type}
\end{align} 
\end{subequations}
where $k$ is Boltzmann's constant, $T$ is the absolute temperature,
$n_i$ is the Si intrinsic carrier density, $N_D$ is the donor concentration,
and $N_A$ is the acceptor concentration.
A parameter can be defined, which is the difference between the Si intrinsic level
and the electrolyte neutral level,
\begin{equation}
\delta_i = 
           \mathrm{E}^\mathrm{Si}_i - \mathrm{E}^\mathrm{Sol}_F =
            \mathrm{E}_i+\Phi^\mathrm{Sol}.
\end{equation}
Using $\delta_i$ and substituting (\ref{eq:wf}) into (\ref{eq:bc}) gives
\begin{subequations}\label{eq:boundary}
\begin{align}
\Delta\psi = V+\frac{\delta_i}{q} + \frac{kT}{q} \ln\left(\frac{N_D}{n_i}\right),
          \ \ \ \ \mathrm{n-type}\\
\Delta\psi = V+\frac{\delta_i}{q} - \frac{kT}{q} \ln\left(\frac{N_A}{n_i}\right),
          \ \ \ \ \mathrm{p-type}.
\end{align}
\end{subequations}
The model uses these as the boundary conditions at ohmic contacts to the Si.

\subsection{Carrier statistics}

The equilibrium charge density in the solution is given by the
Gouy-Chapman theory.\cite{bard,reeves}  For simplicity, the case of a
symmetric electrolyte will be used with the ionic species having
a charge of either plus or minus one.
Using Boltzmann statistics, the net charge concentration
 in the solution is given by
\begin{equation}\label{eq:gouy}
C = C_0 \left(\exp{\left[-\frac{q\psi}{kT}\right]} 
             - \exp{\left[\frac{q\psi}{kT}\right]}\right),
\end{equation}
where $C_0$ is the salt concentration in units of cm$^{-3}$.
$\psi$ and $E_F$ are assumed to be zero in the bulk of the solution.
The first term gives the cation concentration and the second term
gives the anion concentration.

The equilibrium carrier statistics in non-degenerate Si are well known
and can be stated as~\cite{sze}
\begin{subequations}\label{eq:np1}
\begin{align}
n=n_i\exp\left(\frac{E_F-E_i}{kT}\right) \\
 p=n_i\exp\left(\frac{E_i-E_F}{kT}\right).
\end{align}
\end{subequations}
When modeling a semiconductor device composed of a single
material the intrinsic energy level is identified with the
electric potential.\cite{sze,kurata,nussbaum}
In the case of heterojunctions, the discontinuity in the 
intrinsic levels must be taken into
 account.\cite{lundstroma,lundstromb,sutherland}
In the isolated phases, the intrinsic level in the Si is
$E_i=\delta_i$ relative to the Fermi (intrinsic) level of the
solution, Fig.~\ref{fig:energy}a).
When the phases are brought into contact, the Si intrinsic
level is further modified by $\psi$,
and it becomes $E_i = -q\psi +\delta_i$. 
If a voltage is applied to the Si, the Fermi level at the ohmic
contact is raised by the negative of that voltage.
The new Fermi level is $E_F=-qV$.
Putting these values of $E_i$ and  $E_F$ into (\ref{eq:np1}) gives
\begin{subequations}\label{eq:np2}
\begin{align}
n= n_i\exp\left(\frac{q(\psi -\delta_i /q -V)}{kt}\right) \\
p= n_i\exp\left(\frac{q(V+\delta_i /q -\psi)}{kt}\right)
\end{align}
\end{subequations}
The boundary conditions for the electric potential at an ohmic
 contact can obtained assuming electrical neutrality
so that, approximately, $n = N_D$ for n-type Si or 
$p = N_A$ for p-type Si.  Using this in (\ref{eq:np2}) gives
(\ref{eq:boundary}), agreeing with the previous section.

That the Fermi level is constant throughout the Si, with an applied
voltage, can be seen from the following.
In this non-equilibrium case, the Fermi level is replaced by
quasi-Fermi potentials
$E_F\rightarrow -q\phi_n$ for electrons and
$E_F\rightarrow -q\phi_p$ for holes.
The electron and hole currents are proportional to the gradient of the
quasi-potentials.\cite{sze,kurata}
  Due to the insulating SiO$_2$ layer,
the current is essentially zero, so the
quasi-potentials will be constant across the Si
and equal to each other.

\subsection{Diffuse layer structure and surface charges}

The Gouy-Chapman-Stern model will be used to describe the diffuse
layer (or double layer).\cite{bard,reeves}
This model states that there is a thin layer, the Stern layer,
on a surface which contains no salt ions from the solution.
Outside of the Stern layer the ion concentration is given by~(\ref{eq:gouy}).
In this paper, the thickness of the Stern layer will be set to 0.5\,nm.
The capacitance of the Stern layer is then determined by its
dielectric constant.
Experimental results show this capacitance to be about
 20\,\textmu F$\cdot$cm$^{-2}$,\cite{bousse1}
implying a dielectric constant for the Stern layer of
$\epsilon = 1\times 10^{-12}$\,F$\cdot$cm$^{-1}$.

There can also be specifically bound charges at the SiO$_2$ surface
within the Stern layer, as discussed by Sandifer.\cite{sandifer}
To model this a site binding model for hydroxyl groups will be included.
Other specifically adsorbed molecules could be added in a similar way,
for example amine groups.\cite{liu}

The charge on the oxide surface is due to the species MOH, MO$^-$,
and MOH$_2^+$, where M represents Si for a SiO$_2$ surface.
The charge per unit area is given by \cite{landheer,yates}
\begin{equation}\label{eq:site}
\sigma=qN_s\frac{(a^B_\mathrm{H^+}/K_a)\exp(-\beta\psi)
                      -(K_b/a^B_\mathrm{H^+}\exp(\beta\psi)}
                {1+(a^B_\mathrm{H^+}/K_a)\exp(-\beta\psi)
                      +(K_b/a^B_\mathrm{H^+}exp(\beta\psi)}.
\end{equation}
$N_s$ is the site density on the oxide surface,
$\beta = q/kT$,
$a^B_\mathrm{H^+}$ is the activity of protons in the solution bulk,
and the equilibrium constants are given by
\begin{equation}
K_a = \frac{[MOH][a^B_\mathrm{H^+}]}{MOH^+_2]}\,\,\,
\mathrm{and}\,\,\,
K_b = \frac{[MO^-][a^B_\mathrm{H^+}]}{[MOH]}.
\end{equation}
The surface charge will be modeled as a uniformly charged thin layer
with a thickness of
 $t=0.1$\,nm, at the edge of the SiO$_2$,
with a charge density of $\rho = \sigma/t$.

\subsection{Membrane}

A simple membrane model will be used to represent a layer of charged
molecules.\cite{landheer,liu}
It is assumed that the charge is distributed evenly throughout
the membrane.
The salt ion concentration in the membrane is given by the
Boltzmann distribution in the same way as in the Gouy-Chapmann
model for the solution~(\ref{eq:gouy}).
The charge in the membrane is then
\begin{equation}\label{eq:mem}
\rho = qC_m \exp\!\left[\!\frac{q(\delta_m/q-\psi)}{kT}\right] 
     - qC_m \exp\!\left[\!\frac{q(\psi-\delta_m/q)}{kT}\right] + \rho_m.
\end{equation}
$C_m$ is the equilibrium concentration of the salt in the membrane.
It can be expressed by the partition coefficient
$k_s = C_m / C_0$, which gives the ratio of concentration
of each ion crossing the boundary from the solution into the
 membrane.\cite{bard}
$\delta_m$ is the difference between the real potential of a solvated
electron in the membrane and the real potential of a solvated electron
in the solution, analogous to $\delta_i$ in the Si.
If the membrane is composed mainly of water, $\delta_m$ is probably
equal to zero.
$\rho_m$ is the uniform charge density due to the biomolecules.
An example would be a lattice of DNA molecules attached to the
SiO$_2$ by linker molecules as described in Ref.~\onlinecite{poghossian}.
The DNA molecules have a charge of either 1 or 2 electrons
per base unit depending on whether the DNA strand is hybridized
or not.
The linker molecules are assumed to be uncharged and only serve to space
 the membrane a short distance away from the SiO$_2$ surface.

\section{Finite element solution}

Solutions of this model require solving Poisson's equation~(\ref{eq:poisson}),
where the charge density is given for each
region by (\ref{eq:gouy}), (\ref{eq:np2}), (\ref{eq:site}), and
(\ref{eq:mem}).
It would be straight forward to also include charge in the SiO$_2$, if desired.
One advantage of the finite element method is that the exterior
boundary conditions are defined in a natural way and that the
conditions at interior boundaries are matched automatically.
A free, open source finite element solver, Freefem++, \cite{freefem}
was used.
In order to use the finite element method the Poisson equation,
 (\ref{eq:poisson}), must be converted into a variational or weak
 formulation.\cite{freefem,reddy}
While this could be done in three dimensions, the examples in this paper will
be two dimensional.

The full length of a cylindrical nanowire can be
simulated in cylindrical coordinates, in two dimensions using
 radial, $r$ and axial, $z$, coordinates.
The weak form of (\ref{eq:poisson}) is
\begin{equation}
\int \left\{ \frac{\partial v}{\partial r}r\epsilon \frac{\partial\psi}{\partial r}
   + \frac{\partial v}{\partial z}r\epsilon \frac{\partial\psi}{\partial z}
    \right\}\,\mathrm{d}r\mathrm{d}z
  -  \int \left\{ v r q \rho \right\}\,\mathrm{d}r\mathrm{d}z = 0,
\end{equation}
where $v$ is a test function.
As will be shown below, it is often only necessary to simulate the central
cross-section of a nanowire.
In this case cylindrical symmetry is not needed and the 
weak form can be used in $x$-$y$ coordinates 
\begin{equation}
\int \left\{ \frac{\partial v}{\partial x}\epsilon \frac{\partial\psi}{\partial x}
   + \frac{\partial v}{\partial y}\epsilon \frac{\partial\psi}{\partial y}
    \right\}\,\mathrm{d}x\mathrm{d}y
  -  \int \left\{ v q \rho \right\}\,\mathrm{d}x\mathrm{d}y = 0.
\end{equation}
On boundaries with Dirichlet boundary conditions (ohmic contacts) the potential, $\psi$,
must be specified.
For the rest of the boundaries,  Neumann conditions, the above equations
assume that the perpendicular electric field is zero
(by omitting a possible one dimensional integral on the boundary).
This implies that there is zero current crossing the boundary.

Since the charge density is a nonlinear function of $\psi$, the above equations
cannot be solved in a single step. 
A Newton iteration scheme was used \cite{kurata} (see Appendix A).

\section{Calculations}

In this section, calculated results will be given in order to
demonstrate some features of the above model.
Experimentally, it is difficult to determine when the Si is at
equilibrium with the solution.
However, determination of the flat band condition is possible.\cite{bousse1}
The flat band voltage, $V_\mathrm{fb}$, can be obtained from~(\ref{eq:boundary})
by setting $\Delta\psi = 0$,
for example,
\begin{equation}
 V_\mathrm{fb}=-\frac{\delta_i}{q} + \frac{kT}{q} \ln\left(\frac{N_A}{n_i}\right),
          \ \ \ \ \mathrm{p-type}.
\end{equation}
By expressing results relative to the flat band voltage,
accurate values of $\delta_i$ and $E^\mathrm{M}_\mathrm{abs}$
are not needed.

Results shown here will use the following parameters.
The relative dielectric constants of the solution, SiO$_2$, and Si were
78.5, 3.9, and 11.8, respectively.
The temperature was 300\,K and the Si intrinsic carrier density
 was $1.45\times 10^{10}$.
The Si is p-type with a
doping concentration of $1\times 10^{18}$ and
a fixed hole mobility of 130\,cm$^2$V$^{-1}$s$^{-1}$.
The solution concentration was chosen to be 0.01\,M.

\begin{figure}
\centering
\includegraphics[width=3.1in]{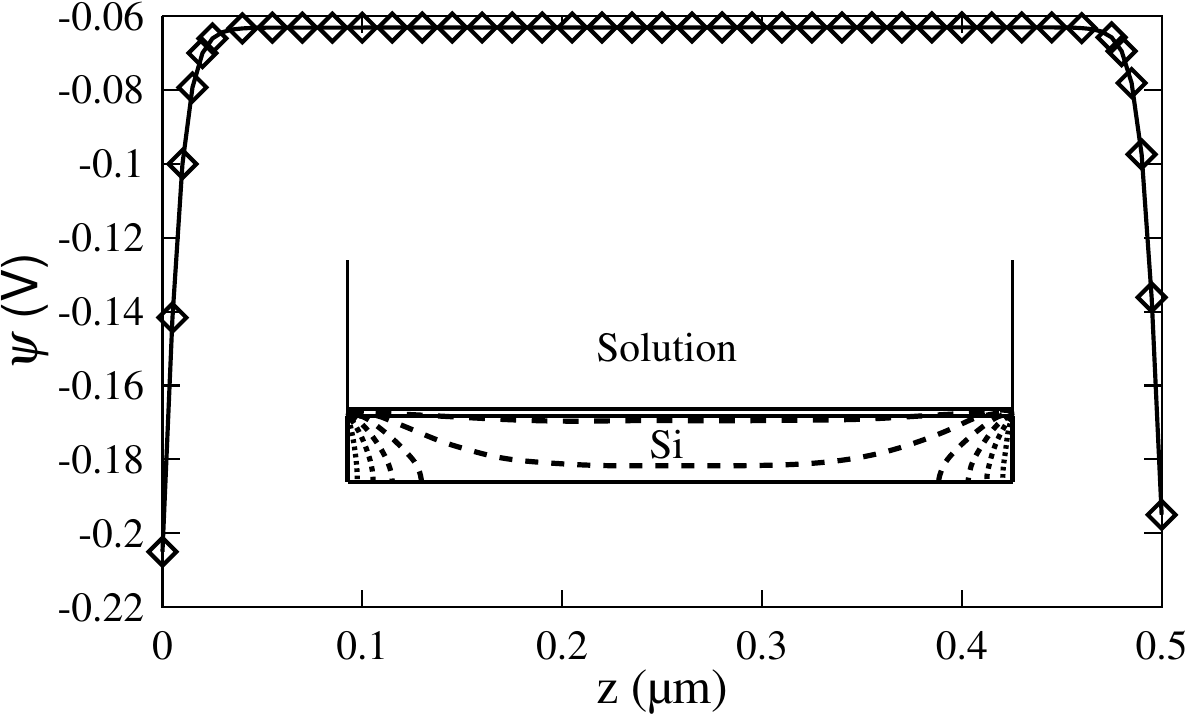}
\caption{Plot of Atlas data of potential along the axis of a 0.5\,\textmu m long nanowire.
The nanowire to solution voltage is $(V_\mathrm{fb}-0.2)$\.V and the drain-source
voltage is $V_\mathrm{ds}=0.01$\,V.
The inset shows a half cross-section
of a 0.1\,\textmu m long nanowire, with constant potential lines.
  The axis of the nanowire
is at the bottom of the drawing.  The source and drain contacts are at the
left and right ends of the Si.  The spacing of the potential contours
 is .03\,V, where the line with the shortest dashes
is -.18\,V and the line with the longest dashes is -.03\,V.
}%
\label{fig:c2zb-16}
\end{figure}

\subsection{Contacts and drain current}

Contact effects and a method to calculate the device response
can be shown with a simple cylindrical model.
This consists of a Si nanowire of radius 10\,nm
 with a 1\,nm thick SiO$_2$ outer layer
in a salt solution with no other charges such as biomolecules
or site binding.
There are ohmic contacts on both ends of the cylinder.
(The Freefem++ code is given in Appendix B.)

The inset of Fig.~\ref{fig:c2zb-16} shows lines of constant potential
when the source and drain voltages are $V=(V_\mathrm{fb}-0.2)$\,V relative
 to the solution.
At this potential the nanowire is in fairly strong depletion.
The structure near the contacts does not change as the nanowire
is made longer.
The end effects due to the contacts extend into the nanowire
a distance of  roughly
twice the radius.
This end effect is independent on the length of the nanowire 
and the results were similar for other gate biases.

While the basic response of a bioFET is the change in carriers in the Si,
a typical experiment will probe this by applying a small drain-source
voltage, $V_\mathrm{ds}$, and measuring the drain current, $I_d$.
A drain-source voltage, $V_\mathrm{ds}$, was applied by setting the drain voltage
(righthand contact in Fig.~\ref{fig:c2zb-16})
to $V+V_\mathrm{ds}/2$ and the source voltage to $V-V_\mathrm{ds}/2$.
When there is a current in the Si the above equilibrium model cannot be used.
Therefore, in this section, current calculations were done with a commercial semiconductor
simulation program, Atlas.\cite{atlas}
An Atlas calculaton of the potential along the axis of the nanowire
 is plotted in
Fig.~\ref{fig:c2zb-16}.
The applied $V_\mathrm{ds}=0.01$\,V can be seen by the difference in $\psi$
 at either end of the nanowire.
One might expect that the potential difference would be distributed evenly
along the length of the nanowire with a slope of 0.02\,V$\cdot$\textmu m$^{-1}$.
However, the Atlas result shows that the potential curve
is nearly flat (slope =$1.67\times 10^{-4}$\,V$\cdot$\textmu m$^{-1}$),
 except near the ends of the nanowire.
This is because the gate (solution) essentially pins the potential distribution
in the nanowire away from the ends.
The current is mainly diffusion current rather than drift current.
It would not be correct to use a gradual channel approximation with a linear
 voltage drop along the nanowire.

\begin{figure}
\centering
\raisebox{1.45in}{(a)} \includegraphics[width=3.1in]{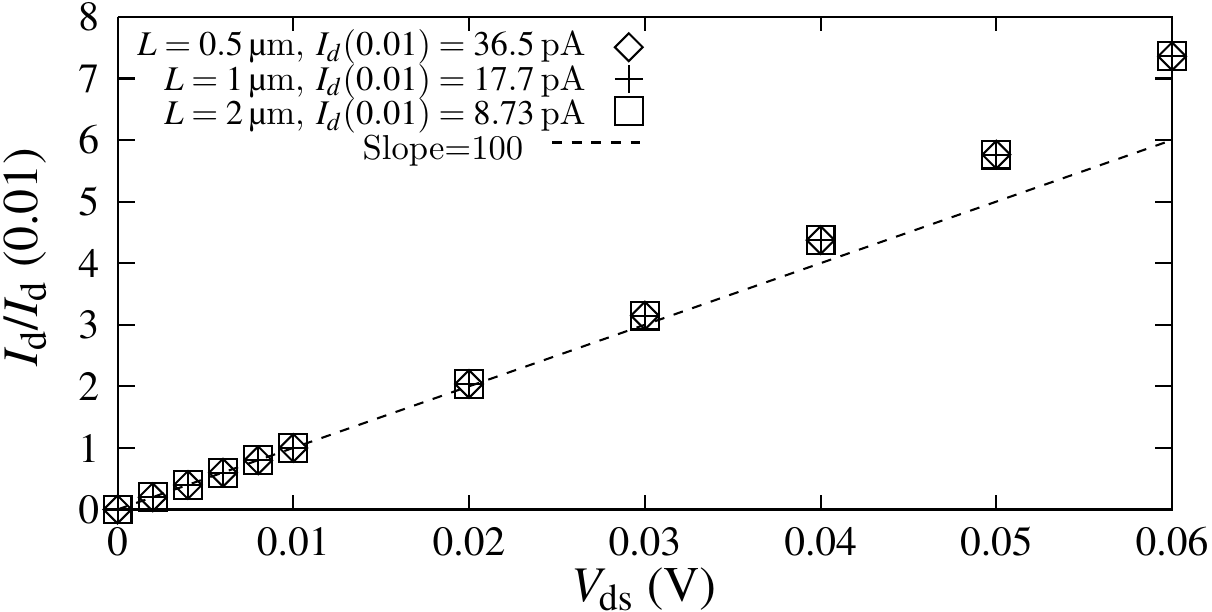}\\[2ex] 
\raisebox{1.25in}{(b)} \includegraphics[width=3.1in]{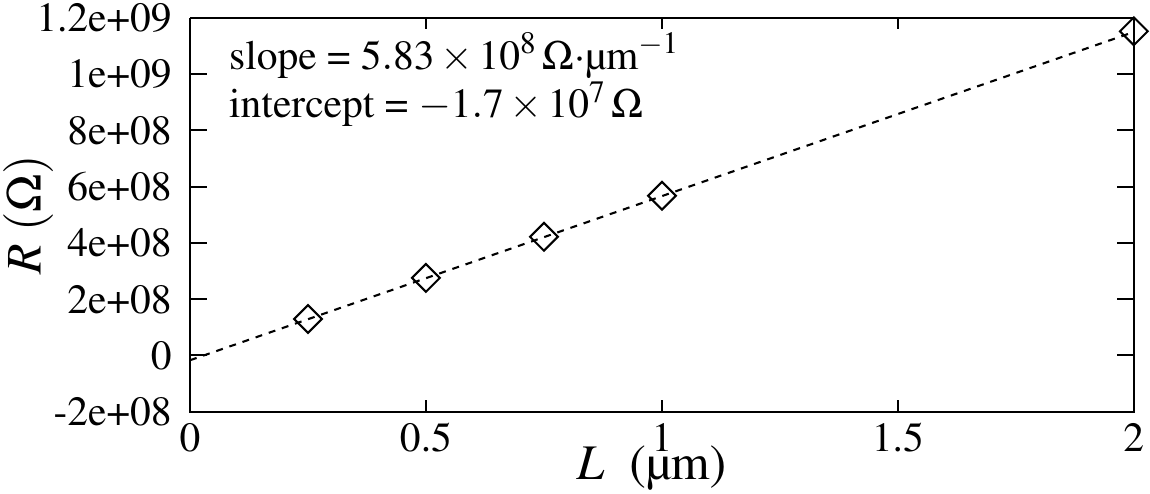} 
\caption{(a) Plot of Atlas data of drain current vs drain-source voltage,
at $V=(V_\mathrm{fb}-0.2)$\,V.
The drain current is normalized by dividing by the value at $V_\mathrm{ds}=0.01$\,V.
The actual current values for $V_\mathrm{ds}=0.01$\,V are given in the key.
(b) Resistance from the Atlas data vs the length of the nanowire.
}%
\label{fig:cy2-Vsd-L}
\end{figure}

The drain current, for three nanowire lengths, is given in Fig.~\ref{fig:cy2-Vsd-L}(a).
When the currents are normalized,
the values for the three lengths agree.
The data is linear up to about 0.02\,V.
The inverse of this slope gives the nanowire resistance, $R$, which is
plotted in Fig.~\ref{fig:cy2-Vsd-L}(b) for a number of lengths.
It is a straight line with a small negative intercept.
$R$ can be accurately modeled as the sum of a correction due to the end resistances
plus the resistance per unit length times the length of the nanowire.

The central cross-section of the nanowire can be simulated,
using the above equilibrium model and Freefem++. A resistance can be
calculated from the average density of holes using $R = L/qp_t\mu$,
where $L$ is the length of the nanowire and $p_t$ is the total number of holes
integrated over the area, per unit length.
It agrees with the resistance found by the Atlas simulations.
This shows that it is only necessary to model the center cross-section of the nanowire.
and that, for small $V_\mathrm{ds}$, the full semiconductor
simulation is not needed.

One would not expect electrons to contribute to a majority carrier hole device
with p-type ohmic contacts; the opposite is sometimes assumed.\cite{liu}
   Atlas calculations confirm that the electrons do not contribute,
even in the case of deep depletion where the number of electrons is similar
to the number of holes.

\subsection{Planar geometry}

One way to make a biofet is to use a Si on insulater (SOI) substrate
and fabricate a ribbon shaped MOSFET.\cite{stern}
A cross-section of this can be modelled.
(The Freefem++ code for this example is given in Appendix C.)
The model assumes a 150\,nm thick burried oxide (SOI wafer) wiht
a 20\,nm thick layer of Si, doped p-type at $1\times10^{18}$\,cm$^{-3}$,
with a 1\,nm oxide layer.
The salt solution has a concentration of $1\times 10^{-2}$\,moles/L.

\begin{figure}
\centering
\includegraphics[width=3.3in]{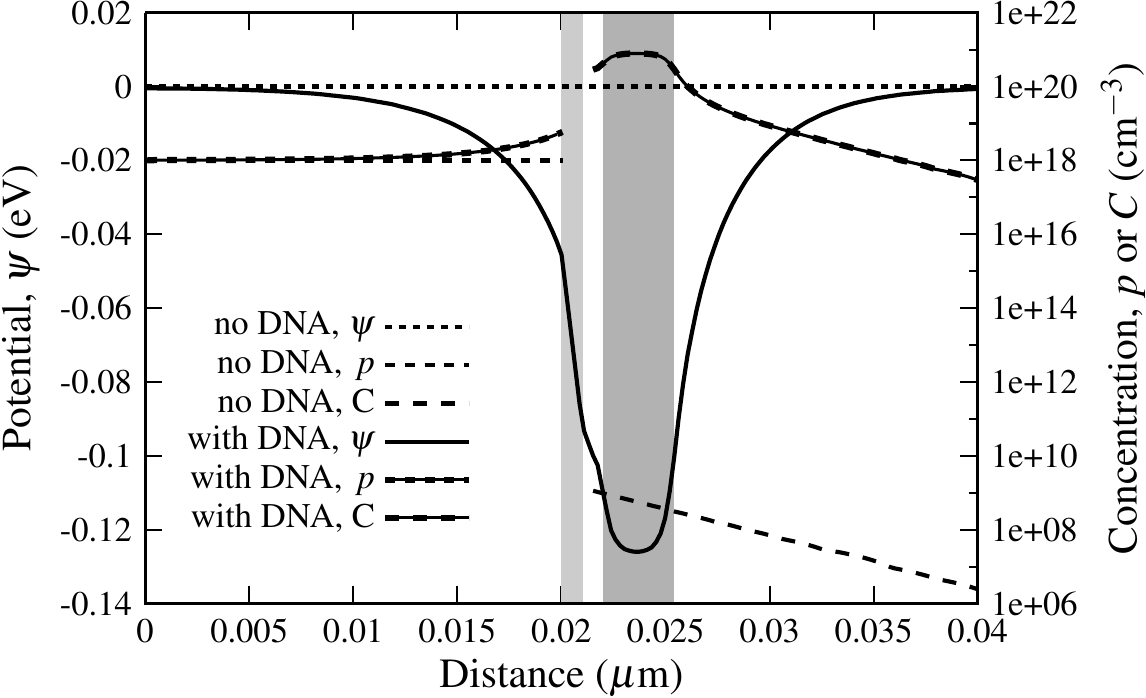}\\[2ex] 
\caption{Plot of parameters at the Si/solution interface of
a ribbon geometry bioFET on a SOI substrate.
The Si layer is on the left, with the lightly shaded band being
the SiO$_2$.
The burried oxide is to the left of the Si and is not shown in
this figure.
To the right of this is solution, including the darker shaded
DNA membrane.
The dashed lined show the potential and hole or net ion 
concentrations at flat band bias with no DNA present.
The other lines show these quantities at the same bias after
a DNA membrane has been added.
}%
\label{fig:p16e-00-02}
\end{figure}

Two situations were modelled. One with no DNA present and with
the Si layer and the substrate bias at the same potential to
give the flat band situation.
Then, with the same bias, a DNA layer was added.
 (The specific values used are
given in the code in the appendix.)
The results are plotted in Fig.~\ref{fig:p16e-00-02}, which
shows only the Si layer and the region of the solution near
the interface.
With no DNA, the potential is flat accross the whole system
and there is essentially no net charge in either the Si or
the solution.
After the DNA is added the solution responds by having a large
excess of positive ions which largely shield the DNA negative
charge.
However, there is still a small response in the Si, seen as the
number of holes above the neutral $1\times 10^{18}$\,cm$^{-3}$
concentration. 
Note that a more realistic simulation would include site binding
charges on the oxide surface.
Code for this is included in the appendix.

\subsection{Sensitivity}

\begin{figure} 
\centering
\includegraphics[width=3.1in]{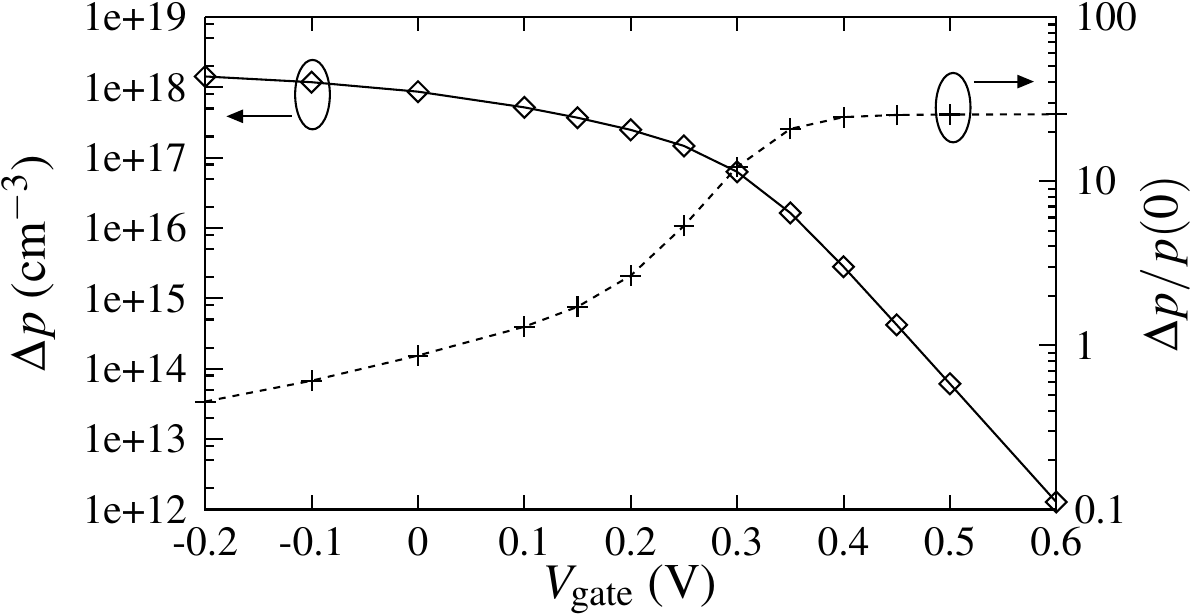} 
\caption{Response, $\Delta p$, vs $V_\mathrm{gate}$ for a circular cross-section
is plotted on the left axis.
Sensitivity, $\Delta p/p(0)$, vs $V_\mathrm{gate}$ 
is plotted on the right axis.
The gate voltage is given by $V_\mathrm{gate}=-(V-V_\mathrm{fb})$. 
}%
\label{fig:sensitivity}
\end{figure}

This section will discuss the response and sensitivity of a nanowire
 to external charge as function of bias.
A nanowire with a circular cross-section surrounded by a charged
membrane was simulated.
Note that there are no contacts to the Si on the central cross-section.
Therefore, it would be wrong to fix the potential at
the center of the nanowire as some workers have done.\cite{li}
The Si nanowire radius was 20\,nm.
The SiO$_2$ layer was 1\,nm thick and the space between this and the
inner surface of the membrane was 1\,nm.
The membrane thickness was 3.4\,nm and its fixed charge density
was $-4\times 10^{20}q$\,cm$^{-3}$.
The change in the number of holes, per unit
 volume, after the membrane is added, $\Delta p$,
is plotted against gate bias 
 in Fig.~\ref{fig:sensitivity}.
For negative values of $V_\mathrm{gate}$ the nanowire is in accumulation.
For positive bias, the nanowire is in depletion and the response drops off exponetially.
This corresponds to the \textit{subthreshold} regime,\cite{gao}
 where the depletion region reaches the center of the nanowire.
Note that this use of the term \textit{subthreshold} is different than the
common usage with respect to an inversion mode FET.\cite{sze}

Some authors define the sensitivity as the response divided by the original
number of holes, $\Delta p/p(0)$, which is also
plotted in Fig.~\ref{fig:sensitivity}.
This sensitivity increases as the depletion deepens and flattens out in
the subthreshold region, in agreement with the approximate, analytic
results of Gao~\cite{gao} and the simulation results of Liu.\cite{liu}
The value of this plateau depends on the doping level, the nanowire radius,
and the concentration of the solution,
as well as the amount of charge in the membrane.
It is important to choose the best gate voltage to obtain optimal properties
of an experimental device.\cite{lu}
Whether $\Delta p$ or $\Delta p/p(0)$ is a better indication of sensitivity
will depend on the specific experimental situation.
Near flat band one would measure a larger absolute change in current,
whereas in the subthreshold region, the current is smaller but
the relative change is larger.


\section{Discussion}

Calculations of other effects on response can easily be done.
Results are not shown here, but generally agree with other publications,
for example the strong effects due to screening of the ions in solution~\cite{nair}
and screening due to the site binding charge.\cite{liu}
In fact, the hole and electron densities,
 the ion concentration, and site binding charge are strongly
interdependent through their dependence on $\psi$, so it is
important to include the entire system in the same calculation.
Also, if a metallic boundary is used,\cite{nair} the calculated response
is much higher than for a semiconducting nanowire.

Cross-sections of nanowire other than circular can be modeled.
For example, calculations show that round nanowires and ribbons
have similar sensitivity.
Some other results of calculations made with this model have been
published.\cite{denhoff}
This includes the response of a trapezoidal nanowire, the effects of back-gating
on a circular nanowire. As well as modeling pH measurements based on the site
binding model, which agreed with experimental measurements.
Note that the salt concentration also
affects the site binding charge.

It was also found that assuming metal boundary conditions for a nanowire
gives much larger sensitivity than for the proper semiconductor boundary
conditions.
This model could also be used to study bioFETs in the inversion mode rather
than the depletion/accumulation mode discussed here.
In this case, the Boltzmann statistics for holes and electrons must be replaced
by the Fermi distribution.
One could also investigate partial coverage of the bioFET by charged
molecules using a three dimensional calculation.

\section{Conclusion}

A semiconductor based model of a bioFET has been developed based on 
the concept of the absolute electrochemical electrode potential.
This model of the response of bioFETs is based on  electrostatic
properties and is valid as long as only small drain-source voltages are used.
The calculation of all the charge distributions is self consistent
and automatically takes account of the boundary conditions at internal
phase boundaries.
This allows it to account for screening due to the ions in a salt solution
and charges specifically bound to the electrode surface.
In this way it accounts for the important physics of the operation
of bioFETs.
The model is built on fundamental principles and can be used as the basis
for more complex and complete models.
It is hoped that this model will supply a basis for the understanding of
experimental results with bioFETs.

\section*{Acknowledgment}

The author would like to thank Dolf Landheer for helpful discussions
about electrochemistry.

\appendix
\section{Newton iteration}
The finite element problem cannot be solved because the source term
is nonlinear in $\psi$.
The approach used here is to start with the weak form of the problem
and apply a Newton iteration to the integrand.
\begin{equation}
  \begin{split}
\int & \left\{ \frac{\partial v}{\partial x}\epsilon \frac{\partial\psi}{\partial x}
   + \frac{\partial v}{\partial y}\epsilon \frac{\partial\psi}{\partial y}
    \right\}\,\mathrm{d}x\mathrm{d}y \\ \nonumber
   &- \int \left\{ v q \left[ n_i \exp(Q(V-\psi)) -N_a \right.\right. \\ \nonumber 
    & + \left.\left. C_0(\exp(-Q\psi)-\exp(Q\psi)\right] \right\}\,\mathrm{d}x\mathrm{d}y = 0.
  \end{split}
\stepcounter{equation} 
\end{equation}
Where $Q=q/kT$, the acceptor concentration is $N_a$, and there are only
  holes in the Si not electrons.
$\psi$ is the potential and $v$ is a test function.
The first half of the second integral (second line) applies only
to the Si and the second half (third line) applies only to
the solution.

Suppose the potential is given by $\psi \,\, \rightarrow w+u$, where
$w$is a (known) guess of the potential and $u$ is a small correction
(to be found). 
Substituting $\psi \,\, \rightarrow w+u$ in the above equation gives
\begin{equation}\label{eq:newton1} \begin{split}
\int & \left\{ \frac{\partial v}{\partial x}\epsilon \frac{\partial w}{\partial x}
   + \frac{\partial v}{\partial y}\epsilon \frac{\partial w}{\partial y}
    \right\}\,\mathrm{d}x\mathrm{d}y
 \\ 
+&\int \left\{ \frac{\partial v}{\partial x}\epsilon \frac{\partial u}{\partial x}
   + \frac{\partial v}{\partial y}\epsilon \frac{\partial u}{\partial y}
    \right\}\,\mathrm{d}x\mathrm{d}y
 \\ \nonumber
 - &\int\left\{vq\left[n_i\exp(Q(V-w))(1-Qu)-N_a+\right]\right\}\mathrm{d}x\mathrm{d}y
 \\ \nonumber
  -&\int\left\{ \left[ C_0(\exp(-Qw)(1-Qu)-\exp(Qw)(1+Qu)\right] \right\}\,
       \mathrm{d}x\mathrm{d}y \\
  & = 0. 
\end{split}
\stepcounter{equation} 
\end{equation}
Where $\exp(-Qu)$ was linearized using a Taylor expansion so that
{$\exp(-Qu)\simeq(1-Qu)$}.
Electrons can be added to the Si using the same method as was used for
 holes in the above.

This can be rearranged to
\begin{equation} \begin{split}
&{
\int \left\{ \frac{\partial v}{\partial x}\epsilon \frac{\partial u}{\partial x}
   + \frac{\partial v}{\partial y}\epsilon \frac{\partial u}{\partial y}
    \right\}\,\mathrm{d}x\mathrm{d}y
} \\ \nonumber
&-\int\left\{-qQn_i\exp(Q(V-w))\,uv\right\}\,\mathrm{d}x\mathrm{d}y
  \\ \nonumber
&-\int\left\{qC_0\left[-Q\exp(-Qw)-Q\exp(Qw)\right]\,uv\right\}\,\mathrm{d}x\mathrm{d}y
  \\ \nonumber
&+\int \left\{ \frac{\partial v}{\partial x}\epsilon \frac{\partial w}{\partial x}
   + \frac{\partial v}{\partial y}\epsilon \frac{\partial w}{\partial y}
    \right\}\,\mathrm{d}x\mathrm{d}y
  \\ \nonumber
&-\int\left\{q(n_i\exp(Q(V-w)) -N_a)\,v\right\}\,\mathrm{d}x\mathrm{d}y
  \\ \nonumber
&-\int\left\{qC_0\left[\exp(-Qw)-\exp(Qw)\right]\,v\right\}\,\mathrm{d}x\mathrm{d}y
= 0. \stepcounter{equation} 
\end{split}\end{equation}
The first three lines are the bilinear terms and the last three lines are linear.
Notice that $-Qn_i\exp(Q(V-w))$ in the second line is just derivative of the
hole concentration $n_i\exp(Q(V-w))$ in the fifth line.
When site binding is added, it can be treated the same way.
The linear term will contain the site binding charge density, $\sigma_0$
and the bilinear term will contain the derivative,
$\partial\sigma_0 / \partial\psi$, which gives
\begin{equation}
  \begin{split}
-\int\left\{\frac{\partial\sigma_0}{\partial\psi}\, uv\right\}\,\mathrm{d}x\mathrm{d}y
  \\ \nonumber
-\int\left\{\sigma_0\,v\right\}\,\mathrm{d}x\mathrm{d}y
\stepcounter{equation} 
  \end{split}
\end{equation}
$\sigma_0$ is a complicated expression.

Solving this finite element problem givesan approximate solution to $u$.
Then a new function $w+u$ is an improved guess for the true potential.
Repeated iteration can further improve this and will tend to converge
towards the true potential solution (depending on a reasonable
first guess). 
\newpage

\onecolumngrid
\section{}
The following is an example input file to Freefem++ of a
model of a cylindrical nanowire.

\begin{lstlisting}
// cyl02.edp 
// cylindrical nano wire  MD  17 Jun 2010
//  uses Newton iteration and includes electrons and holes
//  Stern layer is commented out 
real R = 5.0e-6; // 5e-6 cm = 50 nm
real Rs = 1.0000e-6; // 1.2444e-5; //radius of the Si
real to = 1.0e-7; // silicon oxide thickness
real W = 1.0e-5;
real Q = 38.68;
real q = 1.6e-19;
real ni = 1.45e10;
real es =1.045e-12  ;//1.064e-12; 11.8*8.85418e-14=1.04479324e-12
real ew = 6.9505e-12; // 8.85418e-14 * 78.5,epsilon water
//real eH = 1.0000e-12; // 1.0000e-12= 8.85418e-14 * 11.29, Stern layer
real esio = 3.4531e-13; // 3.9 is epsilon SiO2
real Er = -3.2528; // Si Ei minus Sol real energy, previously I used .2772
real Na = 1e18; // acceptor density, eihter Na or Nd must be zero
real Nd = 0e17; // Donor density
real C0 = 1.0e-2*6.02214e20; // concentration in mol/L* NA / 1000
real D0 = 0e19;
//real aB = 1.0000e-7; // pH = -log_10(aB) = -ln(aB)/2.30258509299404568402
real V = +3.5194300; // potential at ends of Si cylinder, flat band V = +3.7194300
real Vds = 0.0000000; // drain - source voltage, must be small.

border G1(t=0,Rs) {x=t; y=-W; }
border G2(t=Rs,Rs+to) {x=t; y=-W; }
//border G3(t=Rs+to,Rs+to+5e-8) {x=t; y=-W; } // Stern layer
border G4(t=Rs+to,R) {x=t; y=-W; }
border G5(t=-W,W) {x=R; y=t; }
border G6(t=R,Rs+to) {x=t; y=W; }
//border G7(t=Rs+to+5e-8,Rs+to) {x=t; y=W; }
border G8(t=Rs+to,Rs) {x=t; y=W; }
border G9(t=Rs,0) {x=t; y=W; }
border G10(t=W,-W) {x=0; y=t; }

border T1(t=W,-W) {x=Rs; y=t; }
border T2(t=-W,W) {x=Rs+to; y=t; }
//border T3(t=-W,W) {x=Rs+to+5e-8; y=t; }

mesh Th = buildmesh(G1(40)+G2(5)+G4(20)+G5(20)+
                    G6(20)+G8(5)+G9(40)+G10(200)+
                    T1(200)+T2(200));
//mesh Th = buildmesh(G1(20)+G2(5)+G3(1)+G4(2)+G5(20)+G6(20)+
//                    G7(20)+G8(2)+G9(1)+G10(5)+G11(20)+G12(200)+
//                    T1(200)+T2(200)+T3(200)+T4(200));
fespace Ph(Th,P0);
fespace Vh(Th,P1);

Ph reg = region;
plot(reg,fill=1,wait=1,value=1);

int nSi=reg(1e-8,0);
int no=reg(Rs+to/2,0);
//int nH=reg(Rs+to+2.5e-8,0); // Inner Helholtz or Stern layer
int ns=reg(Rs+to+2e-7,0);
cout <<",  nSi "<<nSi<<"  ,no "<< no <<", ns "<<ns<< endl; 

Vh u, v, w, u2;

real cpu=clock();
// Stern layer + eH/q*(region==nH)
Ph epq =  es/q*(region==nSi) + esio/q*(region==no) +
             ew/q*(region==ns);
// The first guess for potenial. Must contain the correct boudary conditions.
 w = ( (V+ Vds*y/W) + Er -(1/Q)*log(Na/ni) )*(sin(pi*y/(2*W)))^2 *(x<Rs); // for p-type
u = 0;

cout << w(1e-6,W) << "  "<< w(1e-6,-W) <<"  "<< w(0,0) <<"  "<< w(R,0) << endl;

problem Lap(u,v,solver=LU) =
      int2d(Th)((dx(u)*epq*dx(v) + dy(u)*epq*dy(v))*x) // Laplace
      +int2d(Th)( ni*Q*( exp(Q*(V+Er-w))+exp(Q*(w-Er-V)) )*u*v*x*(region==nSi) ) // Si
      - int2d(Th)( C0*(-Q*exp(-Q*w)-Q*exp(Q*w))*u*v*x*(region==ns) ) // sol
      +int2d(Th)( (dx(w)*epq*dx(v) + dy(w)*epq*dy(v))*x ) // Laplace
      -int2d(Th)( (ni*( exp(Q*(V+Er-w))-exp(Q*(w-Er-V)) )-Na+Nd)*v*x*(region==nSi) ) // Si
      - int2d(Th)( C0*(exp(-Q*w)-exp(Q*w))*v*x*(region==ns) )  // sol
       + on(G5,u=0) + on(G1,u=0) + on(G9,u=0);

Lap;

cout <<"w+u, w, u:  "<< w(Rs,0)+u(Rs,0) <<"  "<<
       w(Rs+1e-7,0)<<"  "<< u(Rs,0) << endl;

//plot (Th, w, value=true, coef=1, wait=true);
plot (Th, u, value=true, coef=1, wait=true);

u2 = w + u/2;
w = u2;
// Iterate until solution converges; ie. u < about e-16 everywhere
for (int j=1;j<=20;j++){
Lap;
cout << j <<") w+u ,p ,u:  "<< w(Rs,0)+u(Rs,0) <<"  "<<
    ni*exp(Q*(V+Er-w(Rs,0)))<<"  "<<u(Rs,0)<< 
     endl;
u2 = w + u/1;
w = u2;
}

// calculate the total number of holes (per cm length) on the cross-section
// at the middle of the cylinder
int Nn = 80;
real X = 0;
real xdel = Rs/Nn;
real pdel = 0;
real ptot = 0;
for (int n =1;n<=Nn;n++){
X = n*xdel-xdel/2;
pdel = 2*pi*ni*exp(Q*(V+Er-w(X,0)))*(n*xdel-xdel/2)*xdel;
ptot = ptot + pdel;
}
cout <<"total holes "<< ptot <<"  pdel "<< pdel <<"  p_ave "<< ptot/(pi*Rs^2) << endl;

// now calculate the total number of electrons
X=0;
real ndel = 0;
real ntot = 0;
for (int n =1;n<=Nn;n++){
X = n*xdel-xdel/2;
ndel = 2*pi*ni*exp(Q*(w(X,0)-Er-V))*(n*xdel-xdel/2)*xdel;
ntot = ntot + ndel;
}
cout <<"total electrons "<< ntot <<"          n_ave "<< ntot/(pi*Rs^2) << endl;

Vh Ex = -dx(w);
//func Ex = -dx(w);

Ph unitC = 1*(region==ns);
//Ph Cout = C0*(exp(-Q*w)-exp(Q*w))*(region==ns); // This gives digital steps
Ph unitpNa = (region==nSi); // region of the Si trapezoid
//Ph unitf = (region==nf);
X = 0;
real Ni = 400;
{
ofstream file("cy2.dat");
file << "# ave hole density over area of the Si at y = 0,  "<<  ptot/(pi*Rs^2) << "\n";
file << "# ave electron density over area of the Si at y = 0,  "<<  ntot/(pi*Rs^2) << "\n";
for (int i=0;i<=Ni;i++){
X = 0 + i*(0+R)/Ni;
file << X << "  " << w(X,0) << "  " <<
   unitC(X,0)*C0*(exp(-Q*w(X,0))-exp(Q*w(X,0))) <<"  "<<
   unitpNa(X,0)*( ni*exp(Q*(w(X,0)-Er-V)))<<"  "<<
   unitpNa(X,0)*( ni*exp(Q*(V+Er-w(X,0)))) <<
   "  "<< unitpNa(X,0)*(-Na+Nd) <<
   "  "<< Ex(X,0) << "\n";
   }
};

real Y;
{
ofstream file("cyz2.dat");
for (int i=0;i<=Ni;i++){
Y = -W + i*2*W/Ni;
file << Y <<"   "<< w(0,Y)<<"  " << w(Rs/2,Y) << "\n";
   }
}


/*
Vh Ey = -dy(w);
real Nc = 100;
{
ofstream file("nwc2b.dat");
for (int i=0;i<=Nc;i++){
X = 0 + i*Rs/Nc;
file << X <<"  "<< Ey(X,W)<<"  " << Ey(X,-W)<<"  " <<
      q*100*ni*exp(Q*(V+Er-w(X,0)))*Ey(X,W) << "\n";
   }
}
*/
plot (Th, w, value=true, coef=1, wait=true);
plot (Th, u, value=true, coef=1, wait=true);
u2 = w+u;

real[int] viso(6);
//viso[0]=.03;
//viso[1]=0;
//viso[2]=-.03;
//viso[3]=-.06;
//viso[4]=-.09;
for (int i=0;i<6;i++)
  viso[i]=i*-0.03 -0.03;
//for (int i=4;i<9;i++)
//  viso[i]=i*-0.020-0.01;
//for (int i=9;i<viso.n;i++)
//  viso[i]=i*-0.100+0.63;

plot ( u2,viso=viso(0:viso.n-1), value=true,ps="cy-pot.eps", coef=1, wait=true);

\end{lstlisting}

\section{}
The following is an example input file to Freefem++ of a
model of a ribbon nanowire os a SOI substrate.
\begin{lstlisting}
// file "plane-16.edp"  MD 20 Feb 2012
// from planar goemetry with site binding and DNA
//
//  Copyright National Research Council of Canada, 2012.
// This is an input file for Freefem++
// with hydroxyl and amine site binding
// 
real R = 1.2e-5; // Right hand edge of the electrolyte, 5e-6 cm = 50 nm
real L = 1.5e-5; // Left edge of the burried oxide (BOX)
real W = 1.4e-5; // half height of the modellel region
real ts = 2e-6; // thickness of Si layer
real to = 1e-7; // thicknes of SiO2
real tc = 1e-8; // thickness of site binding layer
real tH = 4e-8; // Stern layer thickness is the sum of tc+tH, ie 5 Anstroms
real tL = 5.0e-8; // thickness of linker beyond tH, must be => 1e-8
real tD = 3.40e-7; // thickness of membrane, 3.4e-7 is 10 DNA bases
real Q = 38.68; // Q = q/kT
real q = 1.6e-19; // unit charge
real ni = 1.45e10; // Si intrinsic concentration
real es = 1.045e-12; // dielectric constant of Si, 11.8*8.85418e-14
real ew = 6.9505e-12; // 8.85418e-14 * 78.50,epsilon water
real eH = 1.0000e-12; // 1.0000e-12= 8.85418e-14 * 11.29, Stern layer
real esio = 3.4531e-13; // 3.9 is epsilon SiO2
real em = ew; //1.7708e-12; // membrane dielectric constant
real Na = 1e18; // acceptor density, eihter Na or Nd must be zero
real Nd = 0e17; // Donor density. Nd aded 10 Jun 2010
real C0 = 1.0e-2*6.02214e20; // concentration in mol/L* NA / 1000
real Er = -3.7000; // energy difference Si - electrolyte (I used 0.2772 previously)
real Cm = 1.000*C0; // ion concentration in membrane 
real Em = 0.000; // intrinsic energy level of membrane wrt electrolyte
real cDn = 2; // cDn = 1 for ssDNA, cDn = 2 for dsDNA
real cD = -4.000e20*cDn; //volume charge density of DNA strip, ie. membrane
real Ss; // surface density of site binding charges, calculated below
real Ns = 0;//1.5e14/1e-8; // site binding site density in 0.1 nm thick layer (5.0e14/1e-8)
real Ka = 1e+2;
real Kb = 1e-6;
real Nam = 0;//1.5e14/1e-8; // site binding site density due to amine sites
real Kam = 1e-10; // amine disociation constant
real aB = 1e-7; // pH =7 gives aB = 1e-7
// Vfb = -Er + (1/Q)*ln(Na/ni)
real V = 4.1666266077 - 0.00; // voltage at the drain and source contacts
// for flat band Vback = -(Er + Bwf)
real Vback = +4.2272 - 0.00; // potential at x= -L, assumes the work function of the metal
real Bwf = -0.5272; // Back Si wafer work function = -0.5272 for degenerate p-type
/* Vback is applied to an ohmic contact to Si on the letf edge of the BOX
* The boundary value of the potential must include the workfunction of
* the Si wafer.  If it is degenerate p-type this equals the valance band energy,
* which is -5.25 eV or -0.5272 w.r.t. the Si intrinsic level.  The reference level
* in this model is the work function of the electrolyte and the Si intrinsic level
* is Er w.r.t. that.  So the left boundary condition for potential is
* Vback + Er - 0.5272 or Vback + Er + Bwf which is used in the first guess for w.  */
 

border G1(t=-L,0) {x=t; y=-W; }
border G2(t=0,ts) {x=t; y=-W; }
border G3(t=ts,ts+to) {x=t; y=-W; }
border G4(t=ts+to,ts+to+tc) {x=t; y=-W; }
border G5(t=ts+to+tc,ts+to+tc+tH) {x=t; y=-W; }
border G5b(t=ts+to+tc+tH,ts+to+tc+tH+tL) {x=t; y=-W; }
border G6a(t=ts+to+tc+tH+tL,ts+to+tc+tH+tL+tD) {x=t; y=-W; }
border G6(t=ts+to+tc+tH+tL+tD,ts+to+tc+tH+tL+tD+R) {x=t; y=-W; }
border G7(t=-W,W) {x=ts+to+tc+tH+tL+tD+R; y=t; }
border G8(t=ts+to+tc+tH+tL+tD+R,ts+to+tc+tH+tL+tD) {x=t; y=W; }
border G8b(t=ts+to+tc+tH+tL+tD,ts+to+tc+tH+tL) {x=t; y=W; }
border G9a(t=ts+to+tc+tH+tL,ts+to+tc+tH) {x=t; y=W; }
border G9(t=ts+to+tc+tH,ts+to+tc) {x=t; y=W; }
border G10(t=ts+to+tc,ts+to) {x=t; y=W; }
border G11(t=ts+to,ts) {x=t; y=W; }
border G12(t=ts,0) {x=t; y=W; }
border G13(t=0,-L) {x=t; y=W; }
border G14(t=W,-W) {x=-L; y=t; }

border T1(t=-W,W) {x=0; y=t; }
border T2(t=W,-W) {x=ts; y=t; }
border T3(t=-W,W) {x=ts+to; y=t; }
border T4(t=W,-W) {x=ts+to+tc; y=t; }
border T5(t=-W,W) {x=ts+to+tc+tH; y=t; }
border T6(t=W,-W) {x=ts+to+tc+tH+tL; y=t; }
border T7(t=-W,W) {x=ts+to+tc+tH+tL+tD; y=t; }

mesh Th = buildmesh(G1(40)+G2(20)+G3(4)+G4(4)+G5(3)+G5b(3)+G6a(3)+G6(20)+G7(40)
                   +G8(20)+G8b(3)+G9a(3)+G9(3)+G10(4)+G11(4)+G12(20)+G13(40)+G14(40)
                    +T1(80)+T2(80)+T3(80)+T4(80)+T5(80)+T6(80)+T7(80));
//mesh Th = buildmesh(G1(40)+G2(20)+G3(4)+G4(2)+G5(3)+G6(20)+G7(40)
//                   +G8(20)+G9(3)+G10(2)+G11(4)+G12(20)+G13(40)+G14(40)+
//                    T1(80)+T2(80)+T3(90)+T4(90)+T5(90));

fespace Ph(Th,P0);
fespace Vh(Th,P1);

Ph reg = region;
plot(reg,fill=1,wait=1,value=1);

int nf = reg(-L/2,0); // buried oxide
int nt=reg(ts/2,0);  // Si 
int no=reg(ts+to/2,0); // thin SiO2 on surface of Si
int nc=reg(ts+to+tc/2,0); // charged site binding layer
int nH=reg(ts+to+tc+tH/2,0); // Inner Helholtz or Stern layer
int nL=reg(ts+to+tc+tH+tL/2,0); // solution in linker region
int nm=reg(ts+to+tc+tH+tL+tD/2,0); // DNA membrane
int ns=reg(R-1e-7,0);
cout <<"nf "<<nf<<",  nt "<<nt<<",  no "<<no<<"  ,nc "<<nc<<
      "  ,nH "<<nH<<", nL "<<nL<<", nm "<< nm <<", ns "<<ns<< endl; 

Vh u, v, w;

// The function ep gives the dielectric constant for each region
Ph ep = esio*(region==nf) + es*(region==nt) + esio*(region==no) +
           eH*(region==nc) + eH*(region==nH) + ew*(region==nL) +
            em*(region==nm) + ew*(region==ns);

// This is the first guess for the potential, w.  It must satisfy the boundary conditions.
// That is the potential at the left edge of the BOX must equall Vback
// and the potential at the right edge of the electrolyte must be 0.
w = (Vback+Er+Bwf)*(x/(-L))*(x<0);
//u is the correction (function) in the Newton iteration.
// This problem solves for u, which should get small, about 1e-17 on convergence.
problem Lap(u,v,solver=LU) =
      int2d(Th)((dx(u)*ep*dx(v) + dy(u)*ep*dy(v))) // Laplace
      +int2d(Th)( q*ni*Q*( exp(Q*(V+Er-w))+exp(Q*(w-Er-V)) )*u*v*(region==nt) ) // Si
      - int2d(Th)( q*C0*(-Q*exp(-Q*w)-Q*exp(Q*w))*u*v*(region==ns) ) // sol
      - int2d(Th)( q*C0*(-Q*exp(-Q*w)-Q*exp(Q*w))*u*v*(region==nL) ) // sol in linker
      - int2d(Th)( u*v*q*Ns*(-(aB/Ka)*Q*exp(-Q*w)-(Kb/aB)*Q*exp(Q*w)-4*(Kb/Ka)*Q)/
               (1+(aB/Ka)*exp(-Q*w)+(Kb/aB)*exp(Q*w))^2 *(region==nc) ) // hydroxyl site
      - int2d(Th)( u*v*q*Nam*( -(aB/Kam)*Q*exp(-Q*w) )/
               (1+(aB/Kam)*exp(-Q*w))^2 *(region==nc) ) // amine site
      +int2d(Th)( q*Cm*Q*( exp(Q*(Em-w))+exp(Q*(w-Em)) )*u*v*(region==nm) ) // membrane
      +int2d(Th)( dx(w)*ep*dx(v) + dy(w)*ep*dy(v) ) // Laplace
      -int2d(Th)( q*(ni*( exp(Q*(V+Er-w))-exp(Q*(w-Er-V)) )-Na+Nd)*v*(region==nt) ) // Si
      - int2d(Th)( q*C0*(exp(-Q*w)-exp(Q*w))*v*(region==ns) )  // sol
      - int2d(Th)( q*C0*(exp(-Q*w)-exp(Q*w))*v*(region==nL) )  // sol in linker
      - int2d(Th)( v*q*Ns*(aB/Ka*exp(-Q*w)-Kb/aB*exp(Q*w))/
         (1+aB/Ka*exp(-Q*w)+Kb/aB*exp(Q*w)) *(region==nc) ) // hydroxyl site
      - int2d(Th)( v*q*Nam*(aB/Kam*exp(-Q*w))/
         (1+aB/Kam*exp(-Q*w)) *(region==nc) ) // amine site
      -int2d(Th)( q*(Cm*( exp(Q*(Em-w))-exp(Q*(w-Em)) )+cD)*v*(region==nm) ) // Membrane
      + on(G7,u=0) + on(G14,u=0);
Lap;
cout <<"w+u, w, u:  "<< w(ts,0)+u(ts,0) <<"  "<<
       w(ts,0)<<"  "<< u(ts,0) << endl;

plot (Th, u, value=true, coef=1, wait=true);

w = w + u;
real Xss = ts+to+tc/2; // center of site binding charge at y = 0
real Xsur = ts; // surface of the Si at y = 0
// TODO: write a test for convergence for the following loops
// Right now I manually adjust the number of loops
for (int j=1;j<=11;j++){
Lap;
Ss = Ns*(aB/Ka*exp(-Q*w(Xss,0))-Kb/aB*exp(Q*w(Xss,0)))/
         (1+aB/Ka*exp(-Q*w(Xss,0))+Kb/aB*exp(Q*w(Xss,0)))
        + Nam*(aB/Kam*exp(-Q*w(Xss,0)))/
         (1+aB/Kam*exp(-Q*w(Xss,0))); // sum of hydroxyl and amine charges
cout << j <<") w+u ,p ,u:  "<< w(Xsur,0)+u(Xsur,0) <<"  "<<
    ni*exp(Q*(Er-w(Xsur,0)))<<"  "<<
    u(Xsur,0)<<"   sigma "<< Ss*1e-8 << endl;
w = w + u;
}

// The above iterations take too long to converge, for some values of V and Vback.
// At this point use the solution, w, to generate a new mesh
// with highest grid density in regions where w changes most.
// error= 0.002 controls the grid density and was found by trial and error.
// hmax=2.5e-7 sets the largest grid size.  This is needed to get an accurate enough
// solution in the electrolyte region. Again, the value was found by trial and error.

real error=0.0020;
Th=adaptmesh(Th,w,err=error,hmax=2.5e-7,nbvx=38000);
for (int j=1;j<=28;j++){
Lap;
w = w + u;
Ss = Ns*(aB/Ka*exp(-Q*w(Xss,0))-Kb/aB*exp(Q*w(Xss,0)))/
         (1+aB/Ka*exp(-Q*w(Xss,0))+Kb/aB*exp(Q*w(Xss,0)))
        + Nam*(aB/Kam*exp(-Q*w(Xss,0)))/
         (1+aB/Kam*exp(-Q*w(Xss,0))); // sum of hydroxyl and amine charges
cout << j <<") w ,p ,u:  "<< w(Xsur,0) <<"  "<<
    ni*exp(Q*(V+Er-w(Xsur,0)))<<"  "<<
    u(Xsur,0)<<"   sigma "<< Ss*1e-8 << endl;
}


Vh Ex = -dx(w);
//func Ex = -dx(w);
Ph unitC = 1*(region==ns);
Ph unitL = 1*(region==nL);
Ph unitm = (region==nm); // region of the membrane
//Ph Cout = C0*(exp(-Q*w)-exp(Q*w))*(region==ns); // This gives digital steps
Ph unitpNa = (region==nt); // region of the Si trapezoid
//Ph Dout = w*(region==nt);

// calculate number of total holes, using Freefem++ built in integration
// This is for a 1 cm long trapezoid (nanowire).
real Ptot = int2d(Th)( ni*exp(Q*(V+Er-w))*(region==nt) );
cout <<"total holes integrated over area of Si "<< Ptot << endl;
cout <<"average holes "<<Ptot/(ts*2*W)<< endl;

// calculate number of total electrons, using Freefem++ built in integration
real Ntot = int2d(Th)( ni*exp(Q*(w-V-Er))*(region==nt) );
cout <<"total  electrons integrated over area of Si "<< Ntot << endl;
cout <<"average electrons "<<Ntot/(ts*2*W)<< endl;

// The area of the Si is Area = ts*(2*W)
real Areaint = int2d(Th)( (region==nt) );
cout <<"Area "<< ts*(2*W)<<", AreaInnt "<<Areaint << endl;

cout <<"total surface charges (per cm^2) "<< Ss * 1e-8 << endl;
cout <<"hydroxyl charges (per cm^2) "<< Ns*(aB/Ka*exp(-Q*w(Xss,0))-Kb/aB*exp(Q*w(Xss,0)))/
         (1+aB/Ka*exp(-Q*w(Xss,0))+Kb/aB*exp(Q*w(Xss,0))) * 1e-8 <<
     " ,amine charges (per cm^2) "<<Nam*(aB/Kam*exp(-Q*w(Xss,0)))/
         (1+aB/Kam*exp(-Q*w(Xss,0))) * 1e-8 << endl;

real X;
real Ni = 1350; // number of data points in the file "pl-16.dat"
{
ofstream file("pl-16.dat");
file <<"# Vback = "<< Vback <<",  V = "<< V 
     <<", Area of Si (cm^2)= "<< ts*(2*ts) << "\n";
file <<"# total holes in Si layer  of 1 cm length = "<< Ptot << "\n"; 
file <<"# total electrons in Si layer  of 1 cm length = "<< Ntot << "\n";
file << "# the surface charge (at Xss="<< Xss <<") is (1/cm^2)  " << Ss*1e-8 << "\n";
file << "# Data for the following parameters is given along Y=0."<< "\n";
file <<"# X  potential  electrolyte  membrane  electrons  holes  dopant  electric_field"<<"\n";
for (int i=0;i<=Ni;i++){
X = -L + i*(L+ts+to+tc+tH+tL+R)/Ni;
file << X << "  " << w(X,0) << "  " <<
   unitC(X,0)*C0*(exp(-Q*w(X,0))-exp(Q*w(X,0))) 
     + unitL(X,0)*C0*(exp(-Q*w(X,0))-exp(Q*w(X,0))) <<"  "<<
   unitm(X,0)*( Cm*( -exp(Q*(w(X,0)-Em)) + exp(Q*(Em-w(X,0))) ))<<"  "<<
   unitpNa(X,0)*( ni*exp(Q*(w(X,0)-Er-V)))<<"  "<<  // electron density
   unitpNa(X,0)*( ni*exp(Q*(V+Er-w(X,0)))) <<
   "  "<< unitpNa(X,0)*(-Na+Nd) << // Nd added 10 Jun 2010
   "  "<< Ex(X,0) << "\n";
   }
};

real N;
real Y;
real Tspace = 2e-7; //spacing of data points, 2e-7 = 2 nm
{
ofstream file("tr-ex-pot.dat");
file <<"#  Potential along Si surface"<< "\n";
file <<"# Vback = "<< Vback <<",  V = "<< V <<", site density = "<< Ns << "\n";
file <<"\n\n# X,  Y,  and Potential along the front surface of the Si"<< "\n";
N = (W)/Tspace;
N = 2*round(N);
   for (int i=0;i<=N;i++){
  X = ts;
  Y = (-W) + i*2*(W)/N;
    file << X <<"   "<< Y <<"   "<< w(X,Y) << "\n";
  }
};

plot (Th, u, value=true, coef=1, wait=true);
plot (Th, w, value=true, coef=1, wait=true);
\end{lstlisting}

\end{document}